\documentclass[a4paper]{jpconf}
\usepackage{graphicx}
\begin{document}
\title{The Use of Faraday Rotation Sign Maps as a Diagnostic for Helical Jet Magnetic Fields}

\author{Andrea Reichstein and Denise Gabuzda}

\address{University College Cork, Ireland}

\ead{amn.reichstein@gmail.com, d.gabuzda@ucc.ie}

\begin{abstract}
We present maps of the sign of the Faraday Rotation measure
obtained from multi-frequency radio observations made with the Very Long Baseline
Array (VLBA). The Active Galactic Nuclei (AGN) considered have
B-field structures with a central ``spine'' of B-field orthogonal to the jet
and/or a longitudinal B-field near one or both edges of the jet. This structure
can plausibly be interpreted as being caused by a helical/toroidal jet
magnetic field. Faraday Rotation is a rotation of the plane of polarization
that occurs when the polarized radiation passes through a magnetized plasma.
The sign of the RM is determined by the direction of the line-of-sight B-field
in the region causing the Faraday Rotation, and an ordered toroidal or helical
magnetic field associated with an AGN jet will thus produce a distinctive
bilateral distribution of the RMs across the jet.
We present and discuss RM-sign maps and their possible interpretation regarding
the magnetic field geometries for several sources.
\end{abstract}

\section{Introduction}

\indent When first discovered in the active galactic nucleus (AGN) 1055+018 \cite{1999ApJ...518L..87A}, ``spine--sheath'' magnetic field structure was quite unexpected. It was suggested that, depending on the jet's angle to the line-of-sight, the image might be dominated by the ``spine'' or the ``sheath'', but wouldn't show both at once \cite{1999ApJ...518L..87A}. 
One possible explanation for an AGN with spine--sheath polarization structure is interaction of the jet with the surrounding medium, causing a deceleration of the plasma near the jet edges and stretching out the magnetic field along the jet, forming a shear layer. The polarization is strongest where the interaction is strongest. In the picture proposed in \cite{1999ApJ...518L..87A,1994ApJ...437..122W}, the orthogonal magnetic field along the central axis of the jet is caused by a series of transverse shocks. In the standard Blandford--K\"onigl jet model \cite{1979ApJ...232...34B}, the transverse component of B-field decays less rapidly than the longitudinal field, so that the jet B-field will naturally tend to become more dominated by the toroidal component with distance from the jet base, providing another possible explanation for the spine of orthogonal B-field.

\begin{figure}[hbt]
\center
\includegraphics[width=9cm]{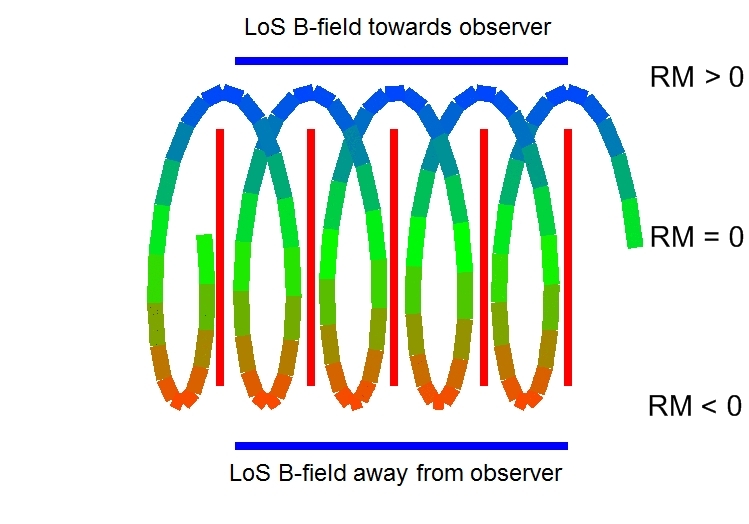}
\caption{Schematic illustration how a helical magnetic field (its projection on the sky) can appear as a spine--sheath structure with the B field orthogonal to the jet axis in the center and longitudinal at the jet edges. In this case, the viewing angle is near 90$^o$ to the jet axis in the jet's rest frame; the rotation measure has opposite signs on either side of the jet, due to the different directions of the line-of-sight B field.}
\label{helix}
\end{figure}

\indent Another possible  explanation of spine--sheath B-field structure is  an overall  helical  magnetic  field around the jet, where the projected toroidal component of the field is dominant at the jet axis, and the longitudinal field component becomes dominant at the  jet edges \cite{2005MNRAS.360..869L,2005MNRAS.356..859P}. This explanation requires only one agent (the helical field), rather than two (the combined presence of a transverse or toroidal field and shear). Figure \ref{helix} shows how to explain a spine--sheath polarization structure with a helical magnetic field. 
To investigate this type of polarization structure, we are making use of the effect of Faraday Rotation.

\section{Faraday Rotation}

\indent When a linearly polarized electromagnetic wave travels through magnetized thermal plasma, the polarization angle $\chi$ rotates due to the different propagating speeds of left and right-circular polarized components of the wave. This is called Faraday Rotation. The amount of rotation is given by:
	\[\chi = \chi_{0} + RM{\lambda}^2\ \ \ \ \ \ \ \ \ \ \ \ \ RM \propto \int{n_{e} B \cdot dl}
\]
where $\chi_{0}$ is the unrotated polarization angle, $n_{e}$ is the electron density, $B \cdot dl$ picks out the line-of-sight magnetic field and $\lambda$ is the observing wavelength.
The coefficient of ${\lambda}^2$ is called the Rotation Measure (RM). Having simultaneous multifrequency observations, the RM can be easily determined by measuring the polarization angle at each wavelength and performing a linear fit on $\chi$ vs. ${\lambda}^2$. This also gives us the intrinsic (unrotated) polarization angle.\newline
\indent Systematic gradients in the RM across the jets of AGN can be interpreted as representing a systematic change in the line-of-sight component of a helical/toroidal B-field (Figure \ref{helix}).
\section{RM-sign maps}

\indent The idea behind RM-sign maps comes from theoretical simulations by Broderick \& McKinney \cite{2010ApJ...725..750B}, who computed sub-parsec-scale B-field and RM distributions of 3D general-relativistic magnetohydrodynamic jets, which they then extrapolated to parsec (observable) scales (the computation time for their simulations to propagate to parsec scales would be prohibitive). They were able to reproduce some of the observed parsec scale RM morphologies, in particular transverse RM gradients produced by the helical field that was generated, and showed that most of the Faraday Rotation occurs in the outer layers of the jet. Even though the observed magnitude of the RM is strongly influenced by finite beam effects, the simulations still show clear gradients when convolved with large beams.
These simulations confirmed that a helical B-field is generated and gives rise to RM gradients across the jet, resulting in a particular bilateral RM structure, which
can be analysed with RM-sign maps. They are also a good tool to help distinguish whether the Faraday screen (the actual Faraday rotating material) is in the immediate vicinity of the jet or is a foreground screen somewhere between the source and the observer. If one is observing a random foreground screen there would be no bilateral structure correlated with the actual jet direction, but some random distribution of Faraday patches.
If the jet is observed side-on in the rest frame of the jet, then the observed RM distribution would ideally be symmetric around a RM of 0 (see Figure \ref{helix}), but in other cases, the RM distribution will be symmetric around some other value.
RM-sign maps can thus be used as an indicator for a helical magnetic field along the jet, and can help exclude random foreground screens.

\section{Observations and results}

\begin{figure}[b]
\center
\includegraphics[width=7.5cm]{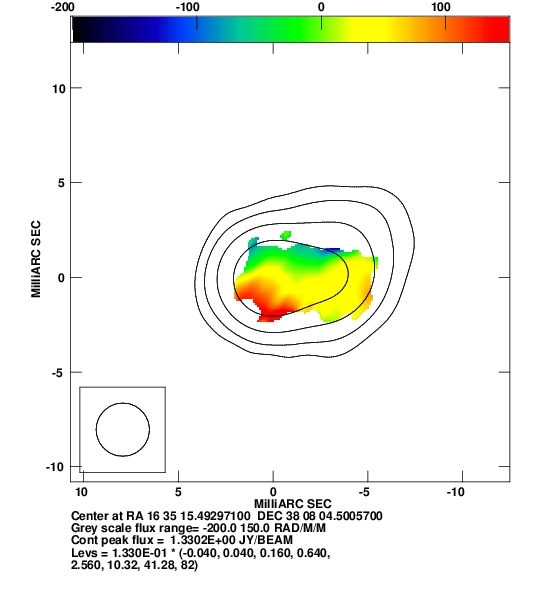}
\includegraphics[width=7.5cm]{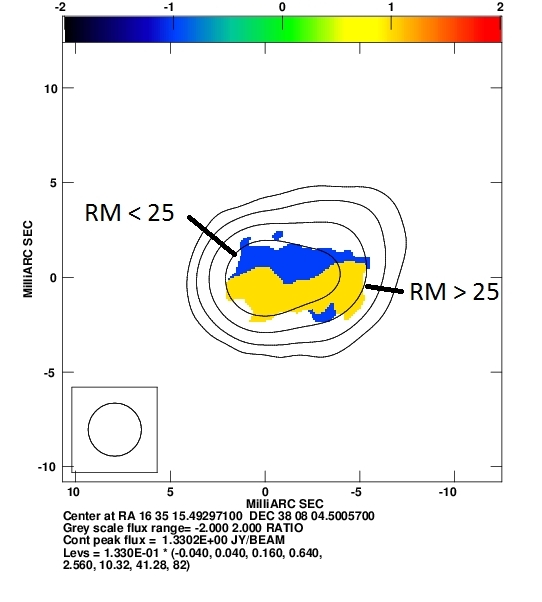}
\label{1633}
\caption{Left: RM map for 1633+382 using six frequencies from 4.6 to 15.4 GHz with contours of the 4.6 GHz total intensity map. Right: RM-sign map for 1633+382. The yellow and blue areas show RMs higher than and lower than 25~rad/m$^{2}$. It shows a bilateral structure correlated with the jet direction.}
\end{figure}

\indent Twelve AGN showing evidence for spine--sheath polarization structures were observed at 4.6, 5.1, 7.9, 8.9, 12.9, and 15.4~GHz with the Very Long Baseline Array (VLBA) on the 27th of September 2007. The data were calibrated and imaged in the NRAO AIPS package using standard techniques. The EVPA calibration was carried out using integrated (VLA) polarization observations of some of the VLBA targets obtained close in time to the VLBA observations. The resulting $\chi$ vs. $\lambda^{2}$ plots are generally consistent with linear behaviour within the $1-2\sigma$ errors. In no case were rotations greater than $45^{\circ}$ observed, consistent with Faraday rotation external to the source volume. Although some of the jets are not well resolved in the transverse direction, it can still be possible to reliably detect the presence of transverse polarization and RM structures, as shown by the simulated transverse profiles of Murphy \& Gabuzda (these proceedings).

\begin{figure}[t]
\center
\includegraphics[width=7.5cm]{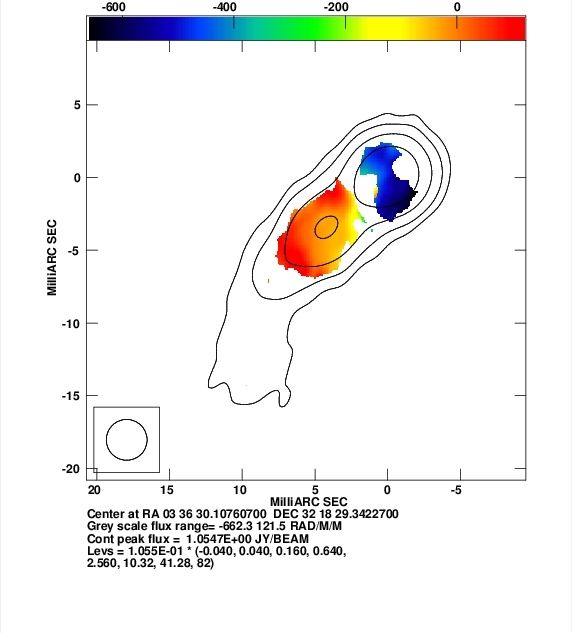}
\includegraphics[width=7.5cm]{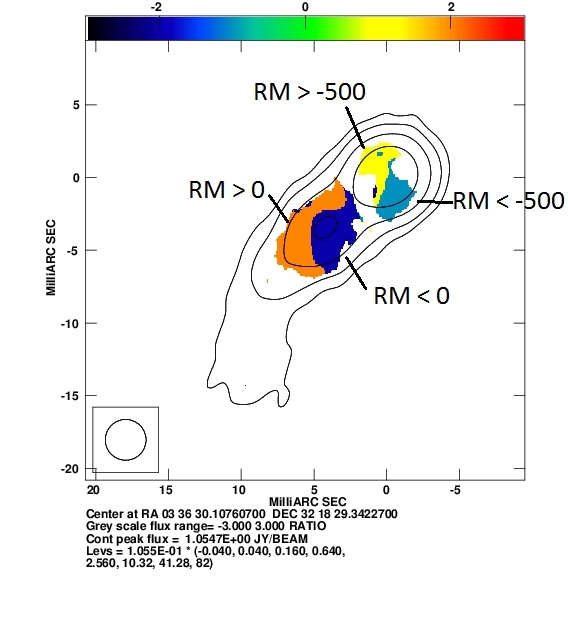}
\caption{Left: RM map for 0333+321 using six frequencies from 4.6 to 15.4 GHz with contours of the 4.6 GHz total intensity map. Right: RM-sign map for 0333+321. The yellow and blue areas in the core indicate RMs higher than and lower than -500~rad/m$^{2}$. In the jet, the orange and dark blue regions show RMs higher than and lower than 0~rad/m$^{2}$. Both parts of the map shows a bilateral structure loosely correlated with the jet direction.}
\label{0333}
\end {figure}

\begin{figure}[htb]
\center
\includegraphics[width=7.5cm]{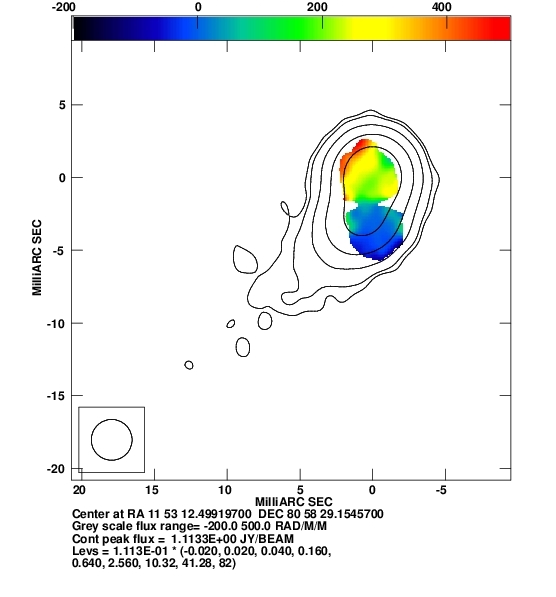}
\includegraphics[width=7.5cm]{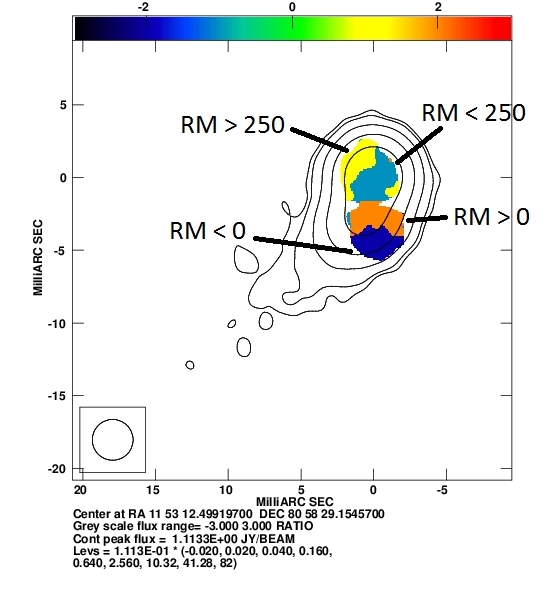}
\caption{Left: RM map for 1150+812 using four frequencies from 4.6 to 8.9 GHz with contours of the 4.6 GHz total intensity map. Right: RM-sign map for 1150+812. The yellow area in the core shows RM higher than 250 $rad/m^{2}$, the blue area RM lower than 250 $rad/m^{2}$. Further out in the jet the orange region indicates RM higher than 0 $rad/m^{2}$, the dark blue area lower than 0 $rad/m^{2}$. The structure in the core region is roughly correlated with the jet direction, however the structure in the jet is not correlated with the jet direction.}
\label{1150}
\end {figure}

\indent Analyzing the RM distributions, we have found some evidence for helical B-field structures in a number of these objects. As an example, Figure 2 shows a transverse RM gradient across the entire resolved jet region of 1633+382. The RM-sign map for this source features a clear bilateral structure correlated with the jet direction,  consistent with a helical/toroidal magnetic field structure surrounding the jet.  
A partially transverse RM gradient was found across the jet of 0333+321, see Figure \ref{0333} and also \cite{2011arXiv1102.0702R}. This result confirms the results of Asada et al. \cite{2008ApJ...682..798A}, who found a very similar RM gradient in this source. The  RM-sign map for this source can be divided into two separate parts, one in the region of the core (roughly symmetrical around an RM of 500~rad/m$^{2}$) and the other one further out in the jet (roughly symmetric around an RM of zero). Even though we are looking at two separate parts of the jet with very different RM values, the direction of the gradient is roughly the same and the RM-sign maps in both parts are roughly correlated with the jet direction. Although there is clearly a transverse component to the RM gradient, the evidence for a helical B-field associated with this jet is weaker than for 1633+382, since the RM gradient is not as close to being transverse.
The different RM values in the core region and jet could be due to somewhat different viewing angles and average electron densities in the core and jet. The RM map of 1150+812 (Figure \ref{1150}) also shows some evidence for transverse gradients \cite{2011arXiv1102.0702R}, however, the RM-sign map shows no consistent evidence for a correlation between the RM structure and the jet direction. The core region seems to show a tentatively transverse structure in the RM-sign map, but the RM-sign pattern in the jet shows no correlation with the jet direction. Thus, we do not find evidence for a bilateral RM-sign distribution indicative of a helical/toroidal jet B-field in this case.

\nopagebreak[4]
\section{Summary}
\indent Faraday Rotation gives information about the line-of-sight B-field. A simple explanation for transverse RM gradients and spine--sheath polarization structures is a helical magnetic field wrapped around the jet, whose changing line-of-sight B-field causes these gradients.
In our images, we have found some evidence for transverse RM gradients in roughly half the analysed sources that show spine--sheath polarization structure \cite{2011arXiv1102.0702R}.
The simulations of Broderick \& McKinney \cite{2010ApJ...725..750B} show that a helical B-field is generated by the jet launching, resulting in a particular bilateral RM structure, which can be analysed using RM-sign maps. Our results so far are varied.
RM-sign maps constructed using our VLBA RM maps show such a bilateral structure for 1633+382, providing supporting evidence that its jet carries a helical B-field. The RM and RM-sign structure in 0333+321 are bilateral, but somewhat oblique, making this a weaker candidate for a jet with a helical B-field. The RM-sign map for 1150+812 does not show a clear and consistent bilateral pattern. 

\indent Overall,  RM-sign maps can potentially provide a useful diagnostic for the presence of helical jet B-fields. However, there is always some probability that RM structures could appear to be correlated with the jet direction purely by chance. This makes it important to try to correlate the presence of bilateral RM-sign structures with other possible indicators of helical jet B-fields, such as spine--sheath polarization structure. The construction of RM-sign maps for larger numbers of AGN jets will also help reveal whether trends indicative of helical B-fields are indeed present in a statistical sense.

\section{References}


\end{document}